\documentclass[twocolumn,showpacs,preprintnumbers,amsmath,amssymb]{revtex4}
\usepackage{epsfig}
\usepackage{amsmath}
\usepackage{bm}

\newcommand{\be}{\begin{equation}}
\newcommand{\ee}{\end{equation}}
\newcommand{\ket}[1]{\left| #1 \right\rangle}

\begin{document}

\title{A Bose-Einstein condensate interferometer with macroscopic arm 
separation}

\author{O. Garcia, B. Deissler, K. J. Hughes, J. M. Reeves and C.A. Sackett}
\affiliation{Physics Department, University of Virginia, Charlottesville, VA 22904}
\email{sackett@virginia.edu}
\date{\today}

\begin{abstract}
A Michelson interferometer using Bose-Einstein condensates
is demonstrated with coherence
times of up to 44 ms and arm separations up to 180~$\mu$m.
This arm separation is larger than that observed for
any previous atom interferometer.
The device uses atoms weakly confined in a magnetic guide
and the atomic motion is controlled using
Bragg interactions with an off-resonant standing wave
laser beam.
\end{abstract}


\maketitle

Atom interferometry, the matter-wave analog of light interferometry, 
works by splitting an atomic wavefunction into two packets that are
separated in space \cite{Berman97,Pritchard01}.  When they are later recombined,
the outcome depends on the difference in their quantum phases. 
Atom interferometry is a powerful measurement tool,
because the phases depend strongly on effects like
inertial forces and electromagnetic fields. 
One limitation, however, has been the difficulty of 
splitting an ensemble of atoms into spatially distinct ``arms.''  
Although individual atomic wave functions can be split over
distances of up to 1.1 mm \cite{Peters01,McGuirk02},
the atoms are typically located randomly
within a cloud or beam that is several mm across.  The separate packets are
therefore not individually accessible in the way that the arms of
an ordinary light interferometer are.

Some applications, such as gravity and rotation measurement, do
not require distinct arms,  and conventional atom interferometry has
proven to be highly effective in these 
cases \cite{Peters01,McGuirk02,Gustavson00,Wicht02}.
However, separated arms permit many additional uses.
In a few atomic beam experiments, separated arms have been 
achieved by using tightly collimated
beams and material diffraction gratings.
This has enabled precise measurements of electric 
polarizability \cite{Ekstrom95,Miffre06}, 
phase shifts in atomic and molecular scattering \cite{Schmiedmayer95},
and atom-surface interactions \cite{Perreault05}.
A larger separation can be expected to have even more utility.

Atoms in a Bose-Einstein condensate are promising for interferometry
due to their low velocities and high spatial 
coherence \cite{Bongs04,Dunningham05}. 
In this case, all the atoms share the same quantum state so the 
arm spacing is the same as the spacing of the individual atomic
wave packets.
Condensate interferometers with packet displacements of over 
100 $\mu$m have been demonstrated \cite{Gupta02,Wang05}, but  
the packets were even larger and did not separate.  
Distinct packets have been obtained by splitting 
a condensate between two optical 
traps \cite{Shin04,Saba05}, but the maximum spacing was only 
13 $\mu$m, comparable to that achieved with beam interferometers.
Recently, a similar spatial separation was obtained 
in a magnetic trap \cite{Ketterle06}.

In our device, $^{87}$Rb condensates are confined in a magnetic
waveguide, as described in Ref.~\cite{Reeves05}.
Condensates with roughly $10^4$ atoms are produced and
loaded into the guide, which is
generated by a set of copper rods mounted in the vacuum chamber.
The guide axis is horizontal and the atoms are held about 1.5 cm from
the rod surfaces.
Ideally, the guide would provide harmonic confinement 
only in the transverse directions, but 
the finite length of our rods leads to axial
confinement as well.  For the data presented here, the 
transverse oscillation frequencies are 3.3 Hz and 6 Hz, 
and the axial frequency is 1.2 Hz.
This confinement is weaker than that of typical magnetic traps,
offering some advantages discussed later.

\begin{figure}
\epsfig{file=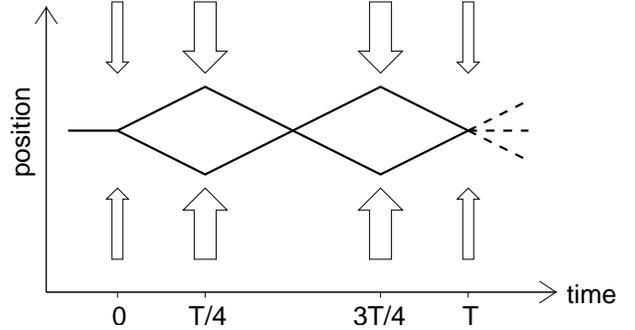,width=3.2in}
\caption{Trajectory of wave packets in the interferometer.
The condensate atoms begin nominally at rest.  At $t=0$, an off-resonant
laser beam (indicated by the arrows)
splits the condensate into two packets traveling
at $\pm 1.2$~cm/s.  At times $T/4$ and $3T/4$, the laser is used to
used to reverse the atoms' motion.
At time $T$, a recombining pulse brings the atoms back to
rest with a probability that depends on the interferometer phase.
The moving atoms (if any) continue to propagate until the system is
imaged to determine the output state.
}
\label{fig:schematic}
\end{figure}

The operation of the interferometer 
is illustrated in Fig.~\ref{fig:schematic}.  
The atoms are 
manipulated by Bragg scattering from an
off-resonant standing-wave laser beam \cite{Martin88,Giltner95,Kozuma99},
with pulses of this beam splitting, 
reflecting, and then recombining the condensate as in Ref.~\cite{Wang05}.
During the splitting pulse, the beam
couples atoms at rest in state $\ket{0}$ to two states
$\ket{\pm v_0}$ moving with speed $v_0 \equiv 2 \hbar k/M = 1.2$~cm/s, 
where $k \approx 2\pi/(780~\text{nm})$ is
the wave vector of the light and $M$ the mass of the atoms.
The coupling is induced by the ac Stark shift, which provides
a potential energy $U = 2\hbar\beta\sin^2 (kz-\alpha)$
with $\beta$ proportional to the light intensity and $\alpha$
denoting the phase of the standing wave pattern.  
Up to an unimportant constant, this can be simplified to
\be
U = \frac{\hbar\beta}{2} \left( e^{-2i\alpha} e^{2ikz} + e^{2i\alpha} e^{-2ikz}\right),
\ee
from which it can be seen that the 
coupling amplitudes are proportional to $e^{\pm 2i\alpha}$.  
For the splitting pulse we take $\alpha =0$ so that 
\be
\ket{0} \rightarrow \ket{+} \equiv
\frac{1}{\sqrt{2}}\big(\ket{+v_0} + \ket{-v_0}\big). 
\ee

After the splitting pulse, the packets freely propagate until 
$t = T/4$, when the Bragg beam is applied so as 
to reverse the packets' direction of motion.  
The packets then propagate for time
$T/2$, crossing each other and separating on the other side.
They are reflected again at $t = 3T/4$ and return to their
initial postion at time $T$.  
If the packets acquire a differential phase $\phi$ during their propagation,
their state is now
\be
\ket{\phi}  = \frac{1}{\sqrt{2}}\left(e^{i\phi/2}\ket{+v_0} 
+ e^{-i\phi/2}\ket{-v_0}\right).
\ee
Since the packets make a full oscillation in the guide,
any phase shift resulting from asymmetry in the potential
cancels to first order, and $\phi$ is nominally zero.
We used this geometry so that the interferometer could be tested with
no uncontrolled phase effects.  

The packets are recombined by applying the same
Bragg pulse used for splitting, but with a variable
standing wave phase $\alpha$.
The pulse therefore couples $\ket{0}$ to the state 
\be
\ket{\alpha} = \frac{1}{\sqrt{2}}\left(e^{2i\alpha}\ket{+v_0} +
e^{-2i\alpha} \ket{-v_0}\right).
\ee
The probability for the recombination pulse to return an atom to
rest at the end of the experiment
is given by the overlap between $\ket{\phi}$ and
$\ket{\alpha}$, 
or $|\langle \alpha \ket{\phi}|^2 = \cos^2(\phi/2 - 2\alpha )$.
After the recombination pulse, the packets are allowed to separate
for 40~ms and the atoms are then imaged.  The fraction of atoms
in the packet at rest is the output signal of the device.

The Bragg beam is derived from a diode laser detuned about 8.4 GHz red
of the $5S_{1/2}$ to $5P_{3/2}$ laser cooling transition.
For the splitting and combining operations, we
obtain good results using a double-pulse sequence
with the theoretically optimum
values of $2^{-5/2} \pi/\omega_r = 24~\mu$s 
for the pulse duration and $\pi/(4\omega_r) = 33~\mu$s 
for the delay between pulses \cite{Wu05}.  Here $\omega_r = \hbar k^2/(2M)$
is the recoil frequency.  The beam power
is 0.7 mW with a Gaussian beam waist of approximately 1.5 mm.

The reflection pulse can be implemented using
second-order Bragg coupling between the $\ket{+v_0}$ and $\ket{-v_0}$ 
states \cite{Giltner95}. However,
this method is very sensitive to velocity errors.
Our condensates sometimes start with a nonzero velocity,
because external magnetic fields can easily disturb the process of
loading the atoms into the waveguide.  We observed this motion 
by taking two pictures of the same cloud using phase contrast imaging.
The residual velocity appears to vary randomly, with a magnitude of
up to $0.5$~mm/s.
In the interferometer, an atom moving with velocity $v_0 + \delta$ will be
reflected to velocity $-v_0 + \delta$, yielding an energy
difference $\Delta E = 2Mv_0 \delta$.  If the reflection pulse
has duration $\tau_r$, then $\Delta E$ must be small compared to
$\hbar/\tau_r$ for the transition to occur.  This requires
$|\delta| \lesssim \hbar/(2Mv_0\tau_r)$.  
For instance, Wang {\em et al.}\ \cite{Wang05} used
$\tau_r = 150~\mu$s, requiring $|\delta| \lesssim 0.2$~mm/s, which is violated
in our experiment.

Increasing the Bragg laser intensity decreases $\tau_r$, but also induces
coupling to the off-resonant $\ket{0}$ state.  We developed a novel
technique that makes use of this coupling.
For $\alpha = 0$, the Bragg beam couples the $\ket{0}$ and $\ket{+}$ states,
making an effective two-level system.
The orthogonal state
$\ket{-} = (\ket{+v_0} - \ket{-v_0})/\sqrt{2}$ is not coupled,
but does acquire a phase shift while the light is on.  
We choose the
pulse intensity and duration so that the atoms make
two full Rabi oscillations between the
$\ket{+}$ and $\ket{0}$ states, ultimately leaving their state unchanged.
During this evolution, atoms in the $\ket{-}$ state acquire a phase of
$\pi$.  This causes $\ket{+v_0} = (\ket{+}+\ket{-})/\sqrt{2}$ to evolve
to $(\ket{+}-\ket{-})/\sqrt{2} = \ket{-v_0}$ and
vice versa, achieving the desired reflection.
The nominal pulse duration is $\pi/(2\omega_r) \approx 67$~$\mu$s
and the amplitude is $\beta = \sqrt{24}\omega_r$, corresponding to
an intensity $\sqrt{3}$ times higher than that of the splitting pulse.
The shorter pulse duration makes this method less sensitive to
the packet velocity, requiring $|\delta| \lesssim 0.5$ mm which
we marginally satisfy.  We observe this pulse to work fairly well,
with reflection efficiencies varying between about 80\% and 100\%.
This variation can presumably be attributed to the fluctuations in
atomic velocity.

To observe the operation of the
interferometer, we vary the recombination phase $\alpha$. 
The Bragg standing wave is generated using a mirror outside the vacuum 
chamber, located a distance $D = 22.5$~cm from the atoms.  
When the laser frequency is changed by $\Delta f$, the 
standing wave phase shifts by $\alpha = (2\pi D/c) \Delta f$.
The frequency change is accomplished in 2 ms by 
adjusting the current of the diode laser.

\begin{figure}
\epsfig{file=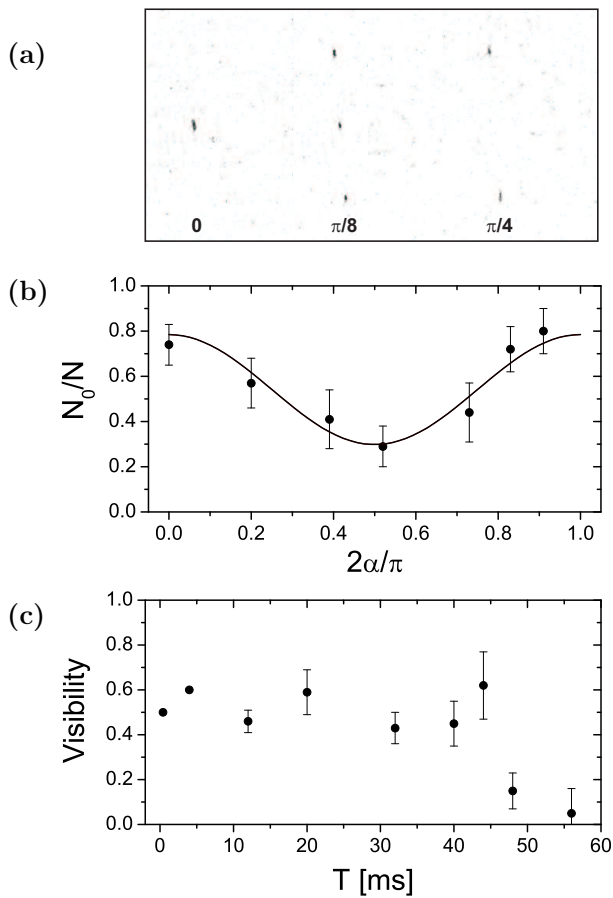,width=3.2in}
\caption{
Interferometer results.
(a) Absorption images of
the interferometer output for 
the indicated standing wave phases $\alpha$.
The dark spots show the positions of the three output wave packets.
These images are analyzed by 
fitting each peak to a Gaussian function to estimate the 
fraction of atoms at rest, $N_0/N$.  Imaging noise introduces
errors of about $\pm 0.05$ to this ratio.
(b) Interference fringe for $T = 40$ ms.  For each value of
the standing wave phase $\alpha$, several images were taken and 
the average value of $N_0/N$ was determined.  
The error bars show the standard deviation of the mean.  
The solid curve is a fit to 
the function $y_0 + A\cos(4\alpha)$ yielding
amplitude $A = 0.24\pm0.05$ and offset $y_0 = 0.54\pm 0.04$.
The visibility $V$ is calculated as $A/y_0 = 0.45 \pm 0.10$.
(c) Visibility as a function of interaction time.
For times $T > 10$~ms, a set of data such as (b)
was acquired and fit to determine the visibility $V$.
For the shorter times, it was not possible to change the laser
wavelength quickly enough, so a set of points at $\alpha = 0$ only
were used to estimate the visibility as $V \approx 2\langle N_0/N \rangle - 1$.
}
\label{fig:results}
\end{figure}

Figure~\ref{fig:results} shows the results of the interference
experiment.  Part (a) shows example images of the spatial distribution
of the atoms for various $\alpha$.
Part (b) plots the fraction of atoms brought to
rest and exhibits the interference fringe.
We performed similar experiments for various values of 
the packet propagation time $T$, finding
the visibility of the interference
to vary as indicated in part (c).  
Interference is observed for $T$ as large as 44 ms.  
Most previous experiments with condensates have been limited
to a coherence time of 10 ms or less \cite{Wang05}, although
a result of 200 ms has been recently reported \cite{Ketterle06}.

We attribute the long coherence time of our experiment
to the weak confinement of our guide.  In particular, 
atomic interactions are much weaker due to the lower density. 
As noted by Olshanii and Dunjko \cite{Olshanii05},
interactions induce a phase gradient on the packets as they
separate, since one end of a
packet stops interacting with the opposing packet immediately
while the other end must traverse the
entire condensate length.
If it takes time $\tau$ for the two packets to fully
separate, the differential phase is on the order of
$\mu\tau/\hbar$ for chemical potential $\mu$.
This can spoil the interference
effect as different parts of the condensate will recombine with different
phases.  Using the Thomas-Fermi approximation \cite{Dalfovo99},
the chemical potential of our initial condensate is
$\mu \approx 2\pi\hbar\times 10$~Hz, 
yielding a phase of about 0.2 rad for $\tau = 3$~ms.
In comparison, the experiment of Wang {\em et al.}~\cite{Wang05} 
had a separation phase of about 3.3 rad.
Weak confinement also reduces
the sensitivity to vibrations of the trap structure 
and requires less precise 
alignment of the Bragg beam to the guide axis \cite{Wu05}.

As seen in Fig.~\ref{fig:results}, we do not observe
perfect interference even for small $T$.  This is due to
run-to-run fluctuations in the interferometer
output that lower the average visibility.
This variation could be attributed to the Bragg beam, but
we monitor the stability of the laser and mirror using an optical
interferometer and observe no significant noise.
The Bragg beam does, however, contain spatial noise that modulates
the beam intensity by about 20\%.  Pointing fluctuations in the
beam therefore change the Bragg coupling strength and causes the
splitting and reflection operations to vary, introducing errors
into the interferometer output.
Another source of noise is the
residual condensate motion mentioned previously, which degrades the
performance of the reflection pulses.

At longer times, the run-to-run fluctuations increase until the
visibility drops to zero.  
The drop appears abrupt, though the error bars are also
consistent with a more gradual decline.  
Various noise sources might cause this,
but it might also stem from the condensate motion.
Another effect of the motion is to
make the splitting pulse asymmetric,
producing more atoms in one packet than the other.  Typical asymmetries
are about 20\%.  The packet with more atoms has a larger self-interaction
energy, leading to a phase shift that fluctuates with the velocity and
increases with $T$.
The observed decoherence time is 
reasonably consistent with this effect, but 
a detailed calculation is difficult because 
the packets evolve in a complex way over time.

\begin{figure}
\epsfig{file=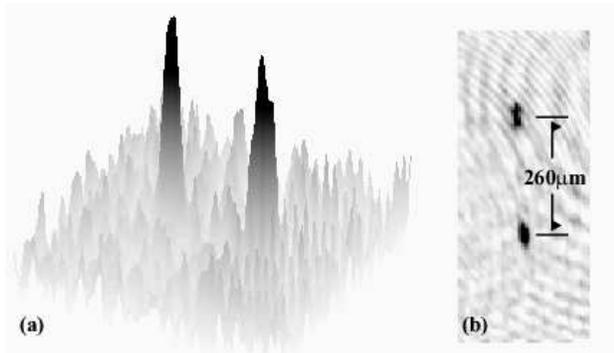,width=3.2in}
\caption{
Separated packets in the interferometer, here shown as 
both a three-dimensional image (a) and a flat picture (b).
The absorption image was taken 11 ms after the splitting pulse and illustrates
the maximum separation for an experiment with $T =44$~ms.
Since the interference visibility is non-zero for this $T$, the
atoms in this picture are in a quantum superposition of 
being in both peaks.
The center-to-center separation of the two peaks is 0.26~mm.
}
\label{fig:split}
\end{figure}

For $T = 44$ ms, the maximum center-to-center packet distance is 
$2v_0 T/4 = 260~\mu$m.
Figure~\ref{fig:split} shows an image taken
at this point.
Each packet has a full width of about 80~$\mu$m, leaving a 180~$\mu$m
spacing between the packets.  To our knowledge, this is the
first literal picture of a matter wave that has been 
split into two demonstrably coherent pieces,
illustrating a fundamental principle of quantum mechanics 
in a concrete way and on a scale that is appreciable to the senses.

This large separation also offers the potential for novel applications.
For instance, one arm of the interferometer might pass into a small
optical cavity, acquiring a phase
shift that depends on the cavity field.  In
this way the number of photons in the cavity could be measured
in a nondestructive way, similar to the
experiments of Nogues {\em et al.}~\cite{Nogues99}.  In comparison,
an interferometer 
would be sensitive to smaller phase shifts, allowing  
the atom-photon interaction
to be non-resonant and making the technique simpler and more flexible.
This could be useful for applications in quantum communication.

Another possibility
would be to have one arm bounce off a material surface 
through quantum reflection \cite{Shimizu01,Pasquini04}. 
This would
allow measurement of the reflection phase shift and provide a
sensitive probe of effects like the Casimir-Polder force.  Although
both arms of our interferometer traverse the same path, 
improvements in the waveguide field stability and homogeneity should 
allow operation with separate paths. 
Alternatively, the motion of the atoms is slow
enough that the surface could be mechanically displaced
before the second packet arrives.

In summary, we have demonstrated a condensate interferometer with
wave packet separations of up to 0.26 mm and clear arm
spacing of up to 0.18 mm.  This is by a large margin the 
greatest arm spacing ever observed in an atom interferometer.
To improve our results, we need to improve the quality of the splitting
and reflecting operations.
We hope to achieve this by using better chamber windows to permit 
a more uniform beam, and by better controlling the net 
magnetic field while loading the guide.
If successful, we estimate that an additional order of
magnitude improvement in packet separation should be possible
before encountering limitations such as phase diffusion \cite{Javanainen97}.
We hope that the techniques
demonstrated here will help condensate interferometers realize
their promise for novel measurement applications.

We are grateful to E.~A. Cornell and R.~R. Jones for 
useful discussions and to K.~L. Baranowski and J.~H.~T. Burke for 
their work on the experiment.
This work was sponsored by the Office of Naval Research 
(Grant No. N00014-02-1-0454) and the National Science Foundation
(Grant No. PHY-0244871).


\end{document}